\documentclass[twocolumn]{aastex631}
\usepackage{CJK}
\usepackage{booktabs}
\usepackage{acronym}
\usepackage{xcolor, soul}
\usepackage{amsmath}
\let\tablenum\relax
\usepackage{siunitx}
\usepackage{threeparttable}

\newcommand{\photoz}{photo-$z$}
\newcommand{\ztrue}{z_{\textrm{true}}}
\newcommand{\zphot}{z_{\textrm{phot}}}
\newcommand{\zspec}{z_{\textrm{spec}}}
\newcommand{\nmad}{\sigma_{\textrm{NMAD}}}
\newacro{GW} {gravitational wave}
\newcommand{\GW}{\ac{GW}}
\newacro{WTS} {Wide-field Time-Domain Survey}
\newcommand{\WTS}{\ac{WTS}}
\newacro{RIS} {Reference Image Survey}
\newcommand{\RIS}{\ac{RIS}}
\newacro{IMS} {Intensive Monitoring Survey}
\newcommand{\IMS}{\ac{IMS}}
\newacro{7DT} {7-Dimensional Telescope}
\newcommand{\sDT}{\ac{7DT}}
\newacro{7DS} {7-Dimensional Sky Survey}
\newcommand{\sDS}{\ac{7DS}}
\newacro{SED} {Spectral Energy Distribution}
\newcommand{\SED}{\ac{SED}}
\newacro{zPDF} {redshift probability distribution function}
\newcommand{\zPDF}{\ac{zPDF}}
\newacro{FWHM}{full width at half maximum}
\newcommand{\FWHM}{\ac{FWHM}}
\newacro{LVF}{Linear Variable Filter}
\newcommand{\LVF}{\ac{LVF}}
\begin{document}
\begin{CJK*}{UTF8}{mj}
\title{Photometric Redshift Forecast for 7-Dimensional Sky Survey}
\author[0000-0002-8130-8044]{Eunhee Ko (고은희)}
\affiliation{SNU Astronomy Research Center, Astronomy Program, Department of Physics and Astronomy, Seoul National University, 1 Gwanak-ro, Gwanak-gu, Seoul 08826, Republic of Korea}
\affiliation{Institut d’Astrophysique de Paris, UMR 7095, CNRS, Sorbonne Universit\'e, 98 bis boulevard Arago, F-75014 Paris, France}
\author[0000-0002-8537-6714]{Myungshin Im}
\affiliation{SNU Astronomy Research Center, Astronomy Program, Department of Physics and Astronomy, Seoul National University, 1 Gwanak-ro, Gwanak-gu, Seoul 08826, Republic of Korea}
\author[0000-0003-3078-2763]{Yujin Yang}
\affiliation{Korea Astronomy and Space Science Institute, Daedeokdae-ro 776, Yuseong-gu Daejeon 34055, Republic of Korea}
\author[0000-0002-1418-3309]{Ji Hoon Kim}
\affiliation{SNU Astronomy Research Center, Astronomy Program, Department of Physics and Astronomy, Seoul National University, 1 Gwanak-ro, Gwanak-gu, Seoul 08826, Republic of Korea}
\author[0000-0001-5342-8906]{Seong-Kook Lee}
\affiliation{SNU Astronomy Research Center, Astronomy Program, Department of Physics and Astronomy, Seoul National University, 1 Gwanak-ro, Gwanak-gu, Seoul 08826, Republic of Korea}
\author[0000-0002-6639-6533]{Gregory S.-H. Paek}
\affiliation{SNU Astronomy Research Center, Astronomy Program, Department of Physics and Astronomy, Seoul National University, 1 Gwanak-ro, Gwanak-gu, Seoul 08826, Republic of Korea}
\correspondingauthor{Myungshin Im}
\email{myungshin.im@gmail.com}
\begin{abstract}
We investigate the expected accuracy of redshifts that can be obtained using low-resolution spectroscopic (medium-band) data from the \sDS\@. By leveraging 40 densely sampled filters with widths of \FWHM\@ = 25~nm, we create \sDS\@ mock catalogs and estimate the redshift accuracy for three \sDS\@ main surveys: \WTS\@, \IMS\@, and \RIS\@. Using photometric redshifts calculated from \texttt{EAZY}, we find that the five-year \WTS\@ provides reliable photometric redshifts with an normalized median absolute deviation ($\nmad$) ranging from 0.003 to 0.007 and a catastrophic failure fraction ($\eta$) from $0.8\%$ to $8.1\%$ at $19 \leq m_{625} < 22$. The spectral resolution $R \sim 50$ of the medium-band dataset effectively captures the 4000~\AA{} break and various emission lines. We also explore the synergy with data obtained from Pan-STARRS1, VIKING, and SPHEREx surveys. Combining the SPHEREx all-sky data with \WTS\@ significantly improves the accuracy of photometric redshift estimates, achieving $\eta = 0.4\%$ and $\nmad = 0.004$ for fainter sources at higher redshifts. The additional near-IR information provided by SPHEREx and VIKING plays an essential role in resolving degeneracies between low and high redshifts. We also observe color excesses by subtracting adjacent broad-band data, which improves the confinement of photometric redshifts and aids in the detection of strong emission line galaxies. 
\end{abstract}
\section{Introduction} \label{sec:introduction}
In extragalactic astronomy, the distribution of galaxies provides crucial insights into the evolution of matter in the universe. A fundamental quantity for measuring their distances is redshift, which reveals how the wavelength of light from an extragalactic source is stretched due to the expansion of the universe. A rich dataset of redshifts allows us to infer extensively the galaxy demographics and related astrophysical phenomena, including star formation (e.g., \citealt{2005ApJ...619L..43A, 2011ApJ...730...61K, 2015ApJ...810...71F, 2021AJ....162...47B}), feedback (e.g., \citealt{2013MNRAS.433.2647S, 2014MNRAS.445..955B, 2023ApJ...948...49R}), and dust attenuation (e.g., \citealt{2010ApJ...719.1250F, 2012A&A...539A..31C, 2013A&A...554A..70B}) over the evolution of the universe. Furthermore, observational cosmology is another testbed that operates with redshifts. In this regard, robust photometric redshift samples provide a comprehensive picture of the universe. In the standard cosmological model, the primary goal is to minimize any uncertainties as much as possible and to resolve the current tension (e.g., \citealt{2011ARA&A..49..409A, 2018arXiv180901669T, 2018ARA&A..56..393M, 2021A&A...650A.148S, 2022PhRvD.105d3512A}).

From a classical point of view, determining the redshift of galaxies involves spectroscopy. Spectral features, such as emission lines and breaks, allow us to calculate the degree of shift in the spectra of galaxies. Although spectroscopic observations provide distance information with high fidelity, they are also time-consuming, expensive, and susceptible to low signal-to-noise ratios (S/N), particularly for faint and distant sources. As an alternative, \textit{photometric redshifts} (\photoz\@) are instrumental in deriving the redshift of a large number of samples based on multi-photometric data. Serving as proxies for low-resolution spectra, photometric redshifts complement the expenses associated with spectroscopic observations (e.g., \citealt{2016ApJS..224...24L, 2021MNRAS.501.6103A, 2022ApJS..258...11W}). Since the concept of photometric redshifts was first suggested in 1960s \citep{1962IAUS...15..390B,1985AJ.....90..418K,1986ApJ...303..154L,2000A&A...363..476B}, we can now derive photometric redshifts in the deeper reaches of the universe with the availability of large-sky survey (\citealt{2019NatAs...3..212S, 2022ARA&A..60..363N} for a recent review). Moreover, they are and will continue to be employed to constrain cosmological probes and investigate astrophysical phenomena in a more statistically robust way. However, despite their advantages, uncertainties of photometric redshifts arising from intrinsically low resolution and contamination still persist as significant challenges.

A novel observational strategy, known as \textit{photo-spectra}, offers a solution to the challenges posed by both low S/N and spectral resolution in previous redshift measurement methods. Functioning as quasi-spectra, the higher spectral resolution achieved through medium- and narrow-band filters enables us to catch the spectral features more precisely, otherwise not resolvable in broad-band filters (e.g., \citealt{2003A&A...401...73W, 2010ApJS..189..270C, 2014arXiv1403.5237B, 2021A&A...654A.101H, 2023arXiv231207581N}). Although it does not reach the full precision of spectroscopic redshifts, this approach can achieve high S/N with relatively short integration time---typically only a few minutes per filter while simultaneously observing thousands of targets within the field of view (e.g., \citealt{2021A&A...653A..31B})---in contrast to spectroscopic surveys, which generally require 1--3 hours per object (e.g., \citealt{2003SPIE.4834..161D}). Unlike spectroscopy, it eliminates the need for a specific target selection process. Furthermore, imaging data, analogous to an integral field unit (IFU) observations, allow the spatially-resolved analysis of nearby galaxies \citep{2014arXiv1403.5237B, 2021A&A...649A..79G}. 

Over the past two decades, a few medium- or narrow-band surveys covering moderately large fields---given the constraints of medium-band surveys, which require more integration time than broad-band surveys---have been conducted. The Classifying Objects by Medium-Band Observations in 17 Filters (COMBO-17, \citealt{2003A&A...401...73W}) survey pioneered the medium-band surveys, whose filter sets provide spectral resolutions of $R$ = 10--20 and 1--2$\%$ of photometric redshift accuracies. With 5 broad-band filters (UBVRI) and 12 medium-band filters, spanning \SIrange[round-precision = 0]{400}{930}{\nm} across three fields over an area of 0.78~$\textrm{deg}^2$, produced reliable photometric redshifts with normalized median absolute deviation of $\nmad\approx 0.03$ for a sample of $\sim 25{,}000$  galaxies. 

Similarly, the Multi-wavelength Survey by Yale-Chile (MUSYC, \citealt{2009ApJS..183..295T, 2010ApJS..189..270C}) is a deep 18-band optical medium-band survey over the $30' \times 30'$ Extended Chandra Deep Field-South. MUSYC combines Subaru medium-band images with existing Subaru $UBVRIzJHK$ from Garching-Bonn Deep Survey (GaBoDS, \citealt{2006A&A...452.1121H}), as well as data from ESO \citep{2003A&A...403..493M}, and SIMPLE \citep{2011ApJ...727....1D}. This integration results in an improvement in redshift accuracy up to $\nmad \sim 0.008$ for the entire sample with $m_{R} < 25.3$. Notably, the addition of deeper medium-band filter data increases the number of fainter sources and reduces the scatter of photometric redshifts. For near-IR observations, the NOAO Extremely Wide-Field Infrared Imager (NEWFIRM, \citealt{2011ApJ...735...86W}) employs near-IR medium-band filter systems covering \SIrange[round-precision = 1]{1}{1.8}{\um}, including $J1, J2, J3, H1, H2$, and $K$ \citep{2009PASP..121....2V}. These filters are designed to probe the Balmer and Lyman break at $1.5 < z < 3.5$. Photometric redshifts from NEWFIRM are in good agreement with spectroscopic redshifts ($\nmad \sim 0.008$ to $\sim 0.017$). 

More recently, the Javalambre-Physics of the Accelerating Universe Astrophysical Survey (J-PAS, \citealt{2009ApJ...692L...5B, 2014arXiv1403.5237B}) aims at observing the Northern sky with 54 narrow-band filters with 145~\AA{} widths covering 3785--9000~\AA{}. For a precursory survey to estimate the scientific feasibility of J-PAS, MiniJ-PAS \citep{2021A&A...653A..31B} demonstrates that a combination of $54$ narrow-band filters and $6$ broadband filters yields highly reliable photometric redshifts ($\nmad < 0.003$) at $m_{r} < 23$ and $odds > 0.82$ \citep{2021A&A...654A.101H}.\footnote{Here, the \textit{odds} parameter refers to the integrated posterior probability within a chosen interval around the best-fit \photoz\@; See \citet{2000ApJ...536..571B} for details.}Additionally, it is shown that further improvement in \photoz\@ calculation is expected with the addition of \SED\@ templates capturing the survey features \citep{2022A&A...668A...8L}. The potential of their narrow-band systems extends from the local universe (e.g., S-PLUS, \citealt{2019MNRAS.489..241M}; J-PLUS, \citealt{2019A&A...622A.176C}) to high-precision \photoz\@ surveys (e.g., PAUS, \citealt{2009ApJ...691..241B, 2023arXiv231207581N}).

The previous efforts of obtaining quasi-spectra with medium- or narrow-bands are not without limitations. The field of view of many of the aforementioned surveys is rather limited at $\ll 1$~deg$^2$. Also, previous attempts utilize a filter wheel or its equivalent, so that simultaneous coverage of all the spectral elements is not possible. Overcoming these weaknesses is important considering the advancement of multi-messenger astronomy and transient surveys, where the discovery and characterization of rapidly evolving transients in a large search area is key to success \citep{2017ApJ...848L..12A, 2021ApJ...916...47K, 2024ApJ...960..113P}.

To overcome the shortcomings of previous works, we have been building the \sDT\@, a multiple telescope system with twenty 0.5~m wide-field telescopes in Chile \citep{2023IAUS..363..207I, 2024SPIE13094E..0XK, Imprep}. By having each wide-field telescope observe at different wavelengths, \sDT\@ enables a nearly simultaneous spectral mapping of the sky over a wide field of view ($\sim 1.2$~deg$^2$) with 40 medium-band filters that cover from 400~nm to 900~nm. The main focus of the project is to rapidly identify the optical counterpart of \GW\@ sources with its spectral mapping capability \citep{2023IAUS..363..207I}. 

Given the power of the all-sky medium-band filters in \sDT\@, we plan to carry out the 7-Dimensional Sky Survey (7DS) and provide spectral maps of the universe at different depths and time cadence to tackle various scientific topics. The \sDS\@ survey aims to unveil various problems spanning multi-messenger astronomy, supermassive black holes, active galactic nuclei (AGN), galaxy evolution, cosmology, the Milky Way galaxy, transients, solar system objects, and exoplanets. The \sDS\@ will be made of three surveys:  \emph{Reference Imaging Survey} (\RIS\@) - all-sky spectral mapping survey to a depth of $\sim 20$~AB mag for a point source detection; \emph{Wide-field Time-Domain Survey} (\WTS\@) - a two-week cadence time-series spectral mapping of a 1{,}000~deg$^2$ area; and \emph{Intensive Monitoring Survey} (\IMS\@) - daily spectral monitoring of about a 10~deg$^2$ field. The stacked images of the time series observations by \IMS\@ and \WTS\@ can reach about 2 and 3 magnitudes deeper than \RIS\@, respectively.

The scientific potential of the \sDS\@ can be maximized when combined with existing or upcoming surveys. For instance, SPHEREx (Spectro-Photometer for the History of the Universe, Epoch of Reionization, and Ices Explorer, \citealt{2014arXiv1412.4872D}) is a space mission to probe the all-sky that can obtain a low-resolution spectral map in the near-IR wavelength range of 0.75--5.0 $\mu$m. Other existing and future missions, such as the Pan-STARRS1 \citep{2016arXiv161205560C}, VISTA surveys \citep{2007Msngr.127...28A, 2013Msngr.154...32E}, and Euclid \citep{2011arXiv1110.3193L}, can provide complementary datasets for \sDS\@. 

In this paper, we evaluate the performance of photometric redshifts of the three different surveys of \sDS\@. Our analysis includes exploring the use of upcoming data and its synergy with other surveys, preparing for the maximization of their combined power. The structure of this paper is as follows. Section~\ref{sec:7DS_in_a_nutshell} outlines the overall background and scientific goals of the \sDS\@'s surveys. In Section~\ref{sec:mock_catalog_generation}, we detail the process of building mock data for the \sDS\@, featuring 40 medium-band filters, and for other systems such as Pan-STARRS1, VIKING, and SPHEREx survey. Forecasts of photometric redshifts obtained from the \sDS\@ and their synergy with other surveys are described in Section~\ref{sec:photoz_prediction_in_7DS} and Section~\ref{sec:synergy_with_other_surveys}, respectively. Section~\ref{sec:discussion} discusses the potential of \sDS\@ for tracing spectral features and highlights further improvements through the inclusion of near-IR information, with particular emphasis on the advantages of its distinctive medium-band system. The conclusion is summarized in Section~\ref{sec:conclusion}. Throughout the paper, the AB magnitude system is adopted \citep{1974ApJS...27...21O}.

\section{\sDS\@ in a nutshell}\label{sec:7DS_in_a_nutshell}
\sDS\@ is a multi-purpose survey designed to perform time-series, wide-field, spatially resolved spectroscopy of the sky, utilizing the nearly simultaneous 40 medium-band imaging capability of the \sDT\@ telescope. \sDT\@ will produce low-resolution spectra for all pixels in the field of view. The two main objectives of the survey are (1) to identify optical counterparts and host galaxies of \GW\@ sources, and (2) to characterize the spectral variability of variable sources in the Universe.

\emph{Reference Imaging Survey}---For the first objective, we will conduct rapid target-of-opportunity spectral mapping of \GW\@ source localization regions. The spectral mapping capability of \sDT\@ integrates wide-field imaging and spectroscopy, allowing for efficient identification of kilonovae. To support this effort, we will carry out the \RIS\@, a spectral imaging survey covering 22{,}000~deg$^{2}$ of the sky at Dec $< +20^{\circ}$. It will reach wavelength-dependent depths of 20~AB mag at 400~nm and 18~AB mag at 875~nm for point-source detection at the 5$\sigma$ level. \RIS\@ images will serve as reference frames to detect new transients and extract their spectra.

In addition, \RIS\@ data will be used to construct galaxy ($z \lesssim 0.5$) and quasar ($z \lesssim 3$) catalogs, which can support statistical inference of host galaxies or active galactic nuclei (AGNs) for \GW\@ sources lacking clear optical counterparts. Our goal is to measure redshifts with an accuracy of 1\% for sources a few tenths of a magnitude brighter than the survey’s limiting magnitude, and a few tenths of a percent when significantly brighter than the magnitude limit. This level of redshift precision is several times better than typical photometric redshifts from broad-band surveys and is necessary for distinguishing galaxy cluster-scale structures without confusion from foreground or background objects (e.g., \citealt{2019ApJ...873..111I,2020A&A...644A..31E,2020AJ....159..258G}).

\emph{Wide-field Time-Domain Survey}---The second component, the \WTS\@, will monitor spectral variability across up to 8,000~deg$^2$ with a cadence of approximately 20~days. A primary target is AGNs, whose long-term ($>$ months) spectral variability can be used to estimate the masses of supermassive black holes via reverberation mapping. The depth of single-epoch \WTS\@ data will match that of \RIS\@, but stacking data over the five-year survey will yield a depth of 22~AB mag (400~nm) for point-source detection. These deep stacked data will facilitate studies of large-scale structures (e.g., galaxy clusters) and galaxy evolution across environments. As with \RIS\@, we require redshift accuracy for galaxies and quasars comparable to that of \RIS\@, but at a fainter magnitude limit. With its wide-area coverage and highly accurate redshifts, \WTS\@ is comparable to the J-PAS survey \citep{2009ApJ...692L...5B, 2014arXiv1403.5237B}, a Stage-IV cosmological survey \citep{Beltrame2024}. We hope to deliver cosmological parameters with comparable precision, if not better, with J-PAS.

\emph{Intensive Monitoring Survey}---Finally, the \IMS\@ is a daily-cadence spectral mapping program of a $\sim$20~deg$^2$ area near the South Ecliptic Pole. Like \WTS\@, its single-epoch depth will match that of \RIS\@, while the five-year stacked depth will reach 23~AB mag (400~nm) for point-source detection. \IMS\@ will track short-term spectral variability of sources such as AGNs, and its stacked data will enable deeper probing of the Universe’s large-scale structure.

\section{Mock Catalog Generation}\label{sec:mock_catalog_generation}
\defcitealias{2003MNRAS.344.1000B}{BC03}

We start the construction of the \sDS\@ mock catalog by using the model \SED\@ catalog from EL-COSMOS \citep{2020MNRAS.494..199S}. EL-COSMOS is a mock \SED\@ catalog based on COSMOS2015 photometric catalog \citep{2016ApJS..224...24L}, covering the COSMOS field uniformly with 31 photometric bands from near ultraviolet to mid-IR (2000~\AA{} to $10^{5}$~\AA{}). This extensive coverage enables the modeling of a homogeneous galaxy population to probe spectral features such as emission lines. As a result, based on the accurate \photoz\@s, verified with spectroscopic redshifts (see \citealt{2016ApJS..224...24L} for further details), the EL-COSMOS provides representative samples of 518{,}404 galaxies, reliably spanning a redshift range of $0 \lesssim z \lesssim 2.5$. 

The \SED\@ modeling process, outlined by  \citet{2020MNRAS.494..199S}, involves building the stellar continuum spectrum based on the same procedure in \citet{2015A&A...579A...2I} and \citet{2016ApJS..224...24L}. They adopted a stellar population synthesis model from \citet[\citetalias{2003MNRAS.344.1000B}]{2003MNRAS.344.1000B} and an initial mass function from \citet{2003PASP..115..763C}. Using the \SED\@ fitting calculation with \texttt{LePhare}, final \SED\@ models are determined as the best-fit template with the minimum $\chi^{2}$. The luminosity and wavelength are corrected by adopting photometric redshifts from COSMOS2015 as the true values. Star-forming nebular regions (continuum emission and discrete emission lines) are added following the approach by \citet{2009A&A...502..423S}.
The number of Lyman continuum photons, $Q_{\textrm{LyC}}$, is obtained from the integrated luminosity of the \citetalias{2003MNRAS.344.1000B} spectra and converted to the luminosity of discrete emission lines ([OII] 3728 \AA{}, $\textrm{H}_{\beta}$ 4863 \AA{}, [OIIIa] 4959 \AA{}, [OIIIb] 5007 \AA{}, and $\textrm{H}_{\alpha}$ 6565 \AA{}). The modeled \SED\@ aligns with the empirical relation between star-formation rate and $\textrm{H}_{\alpha}$ luminosity based on the link between intrinsic emission line luminosities and the number of ionizing photons \citep{2012ARA&A..50..531K}. Furthermore, intrinsic emission line luminosities and dust attenuation are calibrated by taking advantage of matched spectroscopic samples from zCOSMOS-Bright \citep{2007ApJS..172...70L} and 3D-HST \citep{2012ApJS..200...13B, 2016ApJS..225...27M} to match the observed emission lines. On the basis of observations from zCOSMOS-Bright and 3D-HST, a stellar-to-nebular extinction ratio dependent on redshift evolution was derived and applied to the model \SED\@s.
\subsection{\sDS\@ Mock Catalog}\label{subsec:7DS-mock-catalog}
\begin{figure*}[!ht]
    \centering
    \begin{tabular}{c}
    \includegraphics[width=0.7\textwidth]{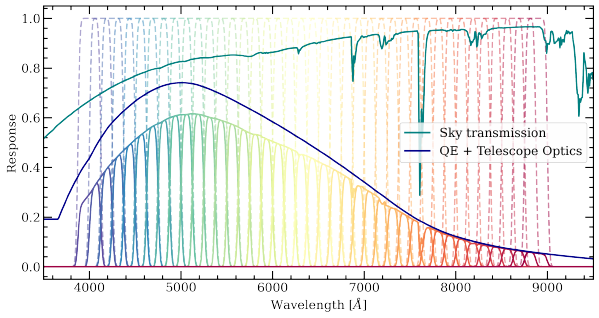}\\
    \includegraphics[width=0.75\textwidth]{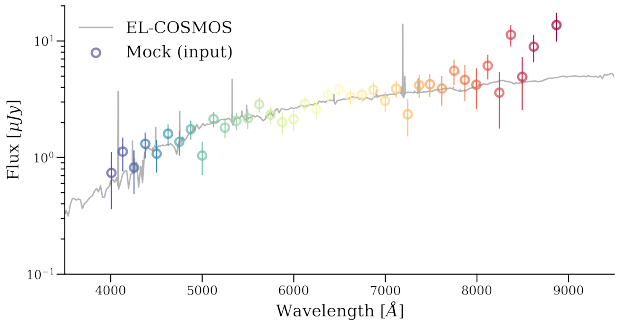}
    \end{tabular}
    \caption{Top panel: Simulated filter transmission curves. Bottom panel: An example of EL-COSMOS \SED\@ (gray line) and its \sDS\@-like mock (data points).}
    \label{fig:filter_transmission_mock_generation}
\end{figure*}
We generate mock transmission curves for the 40 medium-band filters by starting from a smoothed top-hat transmission curve with a \FWHM\@ of 25~nm. To replicate realistic transmission curves as in observation, we incorporate the efficiencies of the telescope, detector, and atmospheric transmission given in Table~\ref{tab:data_specification} \citep{Imprep}. The combined telescope efficiency is convolved with atmospheric transmission,\footnote{\url{https://www.eso.org/sci/software/pipelines/skytools/skycalc}} assuming observational conditions similar to those at the Paranal Observatory, where the Very Large Telescope is located. Given the goal of the prediction study and uncertainties in practical observational conditions, we confirm that this assumption does not affect the results throughout the analysis. The resulting filter transmission curves are presented in the top panel of Figure~\ref{fig:filter_transmission_mock_generation}. Hereafter, we denote the magnitude observed at a filter with a central wavelength of $\lambda$~nm as $m_{\lambda}$ (e.g., $m_{625}$ for the magnitude at the 625~nm filter). 

Note that we do not include the effect of foreground Galactic extinction in our mock photometry for the following reasons: (1) The COSMOS2015 catalog, from which our mocks are built, has already been corrected for Galactic extinction, thereby removing extinction-induced flux biases from our predictions. (2) In practice, the variation of foreground Galactic extinction across the $\sim$2 deg$^2$ COSMOS field is small and difficult to extrapolate to wide-field surveys. The original COSMOS2015 catalog applied extinction corrections using the extinction curve from \citet{Allen1976} and the dust map from \citet{Schlegel1998}, with Galactic reddening $E_{B-V}$ along the line of sight ranging from 0.020 to 0.030 \citep{Galametz2017}. Near the limiting magnitude of the \WTS\@ (see Table~\ref{tab:depths}), this level of $E_{B-V}$ error is subdominant to the 1$\sigma$ photometric error ($\sim \pm0.2$), though it may be non-negligible for brighter samples. Such uncertainties could bias the colors of nearby galaxies, and therefore degrade the overall \photoz\@ accuracy across all redshift ranges \citep{Bordoloi2010,Galametz2017}. To limit the scope of our study as a proof-of-concept, we assume that Galactic extinction does not impact our estimate significantly, while it must be carefully treated in real observations.

Next, we calculate the photometric fluxes detected in the simulated filter windows by using the \texttt{trapezoid} function in \texttt{scipy.integrate}. The wavelength step of the transmission curve is set to 1~nm, which corresponds to the wavelength resolution of EL-COSMOS. This choice ensures accurate flux calculations without missing important spectral features, considering the \FWHM\@ of \sDS\@ medium-band filters (25~nm). Flux errors are determined as a function of exposure time, using the Eq.~\eqref{eq:snr}, with the exposure time set to a five-year observation scenario of \WTS\@ (hereafter \WTS\@ Y5) and assuming 70$\%$ available observation time (180~seconds with 14~days cadence for 5~years and 70$\%$ observation efficiency =15{,}600~seconds). For the \IMS\@ Y5 and \RIS\@, we scaled the exposure time to 156,000 seconds and 600~seconds, respectively, by considering the same condition and different cadence, and using
\begin{equation}\label{eq:snr}
    \text{S/N} = \frac{Q_{\textrm{source}}}{\sqrt{Q_{\textrm{source}} + Q_{\textrm{sky}} + Q_{\textrm{dark}} + Q_{\textrm{readout}}^{2}}},
\end{equation}
where $Q_{\textrm{source}}$ is the number of photons received from the target source during the given exposure time, $Q_{\textrm{sky}}$ from the background sky, $Q_{\textrm{dark}}$ from the dark current of the detector, and $Q_{\textrm{readout}}$ from readout noise.

Lastly, we introduce random noise following a normal distribution with a standard deviation of the true flux error to each flux point. We calculate the 5$\sigma$ depth as the limiting magnitude, at which the S/N from Eq.~\eqref{eq:snr} equals $5$ for each filter set. As a fiducial reference filter, we choose $m_{625}$, whose central wavelength is comparable to the standard broad $R$-band filter and provides robust S/N. The overall 5$\sigma$ depths of the simulated \RIS\@, \WTS\@ Y5, and \IMS\@ Y5 are 20.83, 22.62, and 23.88 respectively at $m_{625}$. Of the 518{,}404 sources in the EL-COSMOS catalog, 20{,}654 galaxies are brighter than the \WTS\@ Y5 depth. 

It is crucial to emphasize that these samples represent the observed data as an ensemble of realizations. An illustrative example of the EL-COSMOS \SED\@ and final photometric mock is depicted in the bottom panel of Figure~\ref{fig:filter_transmission_mock_generation}.
\subsection{SPHEREx Mock Catalog}\label{subsec:SPHEREx-mock-catalog}

\begin{figure*}[!ht]
    \centering
    \includegraphics[width=0.7\textwidth]{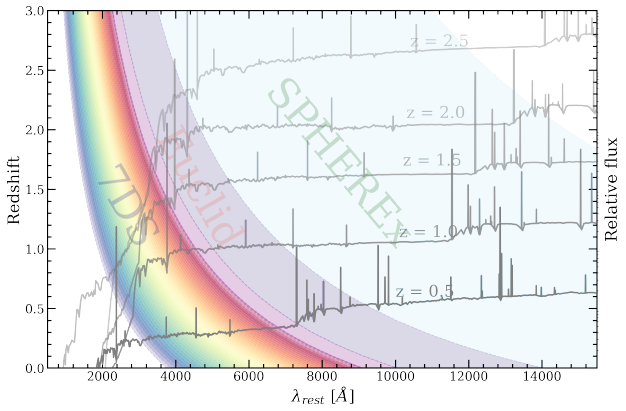}
    \caption{An illustration of the redshifted spectral energy distribution at $z = 0.5, 1.0, 1.5, 2.0, 2.5, 3,0$ and wavelength window filters from \sDS\@ (40 medium-band filters, 4000--8875 \AA{}), Euclid ($J$ and $H$) and SPHEREx (96 \LVF\@s, 0.75--5.0 $\mu$m) as a function of redshifts.}
    \label{fig:redshift_window}
\end{figure*}

One of our primary objectives in leveraging the spectral resolution of medium-bands is to enhance their effectiveness in deriving accurate photometric redshifts with a combination of other complementary surveys. In this context, SPHEREx can bridge the information gap between the optical and near-IR wavelength while maintaining a similar spectral resolution. Figure~\ref{fig:redshift_window} illustrates the redshift window that can be probed by \sDS\@ and other surveys.

The notable feature of SPHEREx is \LVF\@ that is built to shift the pointings in small and consecutive steps. This enables spectroscopic observations without the need for a traditional spectrometer and effective mapping of the sky in terms of both time and field of view. Since the \LVF\@s of SPHEREx capture different central wavelengths as the telescope pointing moves, they are different from fixed bandwidth filters in practice. However, for simplicity, we treat \LVF\@s as 96 photometric filter systems based on previous SPHEREx prediction studies.\footnote{While SPHEREx changed the number of \LVF\@ channels to $102$, we assume that this modification is negligible in the proof-of-concept study as discussed in \citealt{2022ApJ...925..136C}.} We follow the same procedure as in \sDS\@ except for ignoring sky transmission and applying the SPHEREx instrumental parameters from \citet{2014arXiv1412.4872D}. Also, we set the exposure time for the 2 years of SPHEREx all-sky (shallow) survey. The resulting depths range between 19.59 and 17.84 magnitudes over the wavelength range of 0.75 to 5.0 $\mu$~m.\footnote{\url{https://github.com/SPHEREx/Public-products}} For simplicity, we refer to the data from the SPHEREx all-sky survey as ``SPHEREx''.
%
\subsection{Pan-STARRS1 and VIKING Mock Catalogs}\label{subsec:PS1-and-VIKING-mock-catalogs}
We also create mock catalogs for two currently existing surveys: Panoramic Survey Telescope and Rapid Response System (Pan-STARRS1, \citealt{2016arXiv161205560C}) and VISTA Kilo-Degree Infrared Galaxy (VIKING, \citealt{2007Msngr.127...28A, 2013Msngr.154...32E}). This set of choices is to assess the impact of incorporating broad-band optical filters with high S/N and near-IR information. Instead of generating new transmission curves similarly in the \sDS\@ and SPHEREx mock, we use the transmission curves provided by the Pan-STARRS \citep{2012ApJ...750...99T} and VIKING \citep{2007Msngr.127...28A} photometric systems to calculate fluxes. To determine the photometric magnitude error $\sigma_{\rm err}$, we adopt a conversion from $5\sigma$ limiting magnitude $m_{\textrm{lim}}$ and observed magnitude $m_{\textrm{obs}}$ obtained from the calculated flux, as suggested in \citet{2007AJ....134.2236S} and \citet{2019ApJ...873..111I},
\begin{equation}
    \sigma_{\rm err}^{2} = (0.04 - \gamma)x + \gamma x^{2},
\end{equation}
where $x = 10^{0.4(m_{\text{obs}}-m_{\text{lim}})}$ and $\gamma = 0.04$. Here $\gamma$ refers to a factor dependent on possible noise sources, such as sky brightness and/or readout noises. By setting $\gamma = 0.04$, we assume that these effects are properly processed with little systematic errors after calibration (e.g., \citealt{2018AJ....155....1G, 2020AJ....159..258G, 2020A&A...644A..31E}). 

We adopt Pan-STARRS1 (PS1) $m_{\textrm{lim}}$ values ($g, r, i, z, y$; AB magnitude) as (23.3, 23.2, 23.1, 22.3, 21.4) from \citet{2016arXiv161205560C}. The photometric system of VIKING is originally given in Vega magnitude, so we convert its $m_{\text{lim}}$ values ($z, Y, J, H, K_{s}$) described in \citet{2021MNRAS.501.1663B} to AB counterparts (22.1, 20.3, 20.9, 19.8, 19.2).\footnote{\url{http://casu.ast.cam.ac.uk/surveys-projects/vista/technical/filter-set/}} Finally, we introduce random error to the true flux, same as in the \sDS\@ and SPHEREx mock. 

We summarize the overall survey depths used in our study in Table~\ref{tab:depths}.

\startlongtable
\begin{deluxetable*}{ccc}
\label{tab:depths}
\tabletypesize{\scriptsize}
\tablewidth{0pt}
\tablecaption{5$\sigma$ magnitude depth in 7DS, PS1, VIKING, and SPHEREx \label{tab:depth}}
\tablehead{
  \colhead{Surveys} & \colhead{Filter} & \colhead{5$\sigma$ limiting magnitude}
}
\startdata
    \sDS\@ & & \begin{tabular}{ccc}  \RIS\@ &\WTS\@ Y5 & \IMS\@ Y5 \end{tabular}\tablenotemark{\scriptsize a}\\
     & $m_{400}$ &  \begin{tabular}{ccc}  \num[round-precision = 2]{21.17414022} & \num[round-precision = 2]{22.98861824} & \num[round-precision = 2]{24.24033981} \end{tabular}\\
    & $m_{413}$ & \begin{tabular}{ccc}  \num[round-precision = 2]{21.21873636} & \num[round-precision = 2]{23.02257071} & \num[round-precision = 2]{24.27387831} \end{tabular}\\
    & $m_{425}$ & \begin{tabular}{ccc}  \num[round-precision = 2]{21.31344035} & \num[round-precision = 2]{23.11481268} & \num[round-precision = 2]{24.36602572} \end{tabular}\\
    & $m_{438}$ & \begin{tabular}{ccc}  \num[round-precision = 2]{21.33358231} & \num[round-precision = 2]{23.13093603} & \num[round-precision = 2]{24.38199567} \end{tabular}\\
    \midrule
    Pan-STARRS1 & $g$ & 23.3\tablenotemark{\scriptsize b}\\
       & $r$ & 23.2 \\
       & $i$ & 23.1 \\
       & $z$ & 22.3 \\
       & $y$ & 21.4 \\
    \midrule
    VIKING & $z$ & 22.1\tablenotemark{\scriptsize b}\\
     & $Y$ & 20.3\\
     & $J$ & 20.9\\
     & $H$ & 19.8\\
     & $K_{s}$ & 19.2\\     
    \midrule
    SPHEREx & \num{7.500000e-01}\tablenotemark{\scriptsize c} & \num{1.958756e+01}\tablenotemark{\scriptsize a}\\
    & \num[round-precision = 2]{7.682927e-01} & \num[round-precision = 2]{1.957540e+01}\\
    & \num[round-precision = 2]{7.870315e-01} &  \num[round-precision = 2]{1.956413e+01} \\
    & \num[round-precision = 2]{8.062274e-01} & \num[round-precision = 2]{1.955372e+01}\\
    & \num[round-precision = 2]{8.258915e-01} & \num[round-precision = 2]{1.954414e+01} \\
\enddata
\tablenotetext{a}{5$\sigma$ depth in $m_{\text{AB}}$ at which S/N $= 5$ computed on the customized filter curves for the point source detection (see Section~\ref{subsec:7DS-mock-catalog}--\ref{subsec:SPHEREx-mock-catalog}).
}
\tablenotetext{b}{Same 5$\sigma$ depth as (a), but computed on the actual filter curve (see Section~\ref{subsec:PS1-and-VIKING-mock-catalogs}).
}
\tablenotetext{c}{Central wavelength in $\mu$m. Note that SPHEREx is not filter-based but \LVF\@s system.
}
\tablecomments{Only a portion of this table is shown here. The full version of the table is available in the online supplementary material.} 
\end{deluxetable*}
\subsection{Photometric Redshift: EAZY}
We use the \texttt{EAZY} code for calculating photometric redshifts \citep{2008ApJ...686.1503B}. \texttt{EAZY} is an optimized tool that employs template fitting and a Bayesian approach to determine photometric redshifts efficiently. The templates from \texttt{EAZY} are constructed based on the PEGASE models \citep{FiocRoccaVolmerange1997}, which comprise $\sim$~3000 spectra spanning ages from 1~Myr to 20 Gyr with diverse star formation histories. In this work, we use the \texttt{eazy\_v1.3} template set, where emission lines are added following the prescription of \citet{2009ApJ...690.1236I}. 

From these input templates, the \textit{nonnegative matrix factorization} (NMF) algorithm \citep{2007AJ....133..734B} can find an optimized group of templates and template error function, reducing calculation time in the absence of spectroscopic samples (see \citealt{2008ApJ...686.1503B} for more details). This feature aligns with our study's purpose of probing the photometric information through iterative experiments. However, considering the survey's wide-ranging data and advancing technologies, maximizing its potential is crucial. We leave future improvements to obtain more accurate photometric redshifts for our future study.

Among various point estimates of $\zphot$ provided by \texttt{EAZY}, we adopt \texttt{z\_a} unless otherwise specified (e.g., application of priors). This estimate corresponds to the redshift that minimizes $\chi^{2}$ from linear combinations of all templates, without applying any prior. In our tests for the non-prior case, \texttt{z\_a} yields better \photoz\@ performance than \texttt{z\_{m1}}, which marginalizes over the Gaussian likelihood function $\exp{(-\chi^{2})}$. Moreover, \texttt{z\_a} directly links the photometric measurements to their best-fitting templates, making it a natural choice for our analysis.

To explore the potential for future improvements, we further investigate improvements by applying an apparent magnitude prior. For the results obtained with the inclusion of priors, we adopt \texttt{z\_m2}, which has been shown to provide the best \photoz\@ estimate with priors \citep{2008ApJ...686.1503B}. Following \citet{2000ApJ...536..571B}, the apparent magnitude prior $p(z|m_{625})$ is customized by fitting the redshift distribution of galaxies within a given magnitude bin $m_{625, i}$ to the functional form:
\begin{equation}\label{eq:prior}
    p(z|m_{625, i}) \propto z^{\gamma_{i}} \exp{\left[ -(z/z_{0,i})^{\gamma_{i}} \right]},
\end{equation}
where $\gamma_{i}$ and $z_{0,i}$ are fitting parameters and $i$ denotes the $i$-th galaxy in the catalog. The priors are determined by dividing the EL-COSMOS sample into $m_{625}$ magnitude ranges from $15$ to $30$ with intervals of $0.5$, and fitting them into Eq.~\eqref{eq:prior} for each bin. This functional form assigns a low probability weight to bright galaxies at high redshifts. 

Constructing priors based on the true redshifts in EL-COSMOS is a reasonable choice, as it minimizes the risk of adopting inappropriate priors: the reference catalog (COSMOS2015) is both deeper and more complete than our target survey \sDS\@ \citep{2016ApJS..224...24L,2018MNRAS.478..592H}. Note that propagated uncertainties in these priors---arising from Poisson sampling errors (shot noise) and/or sample variances due to large-scale structures---are expected to append further uncertainties at the 1$\%$ level for $z < 1.5$. However, theoretical approaches to mitigate these effects already exist (e.g., \citealt{2020MNRAS.498.2984S}). Given the reliability of our reference data, we suppose the additional uncertainties from our customized priors are ignorable.

In practical applications, priors are expected to be constructed from a large spectroscopic (e.g., \citealt{2016arXiv161100036D}) and complementary photometric samples (e.g., \citealt{2022ApJS..258...11W,2022MNRAS.512.3662D}) that offer high completeness at $z \lesssim 1$. Looking ahead, upcoming datasets such as LSST \citep{2018AJ....155....1G} and Euclid \citep{2020A&A...644A..31E} will also serve as prospective resources for building reliable priors at higher redshifts.

For characterizing the accuracy of photometric redshifts, we introduce three metrics: \textit{Catastrophic failure} $\eta$, \textit{NMAD} $\nmad$, and \textit{Bias} $b$.
\begin{itemize}
    \item \textit{Catastrophic failure}, $\eta$, is the fraction of severely deviated photometric redshifts from spectroscopic redshifts more than $15\%$:
    $$\eta = \rm fraction(|\zspec - \zphot| / (1 + \zspec) > 0.15).$$
\end{itemize}
\begin{itemize}
    \item \textit{NMAD} (Normalized Median Absolute Deviation), $\nmad$, represents the $1\sigma$ uncertainty of photometric redshifts in a given spectroscopic redshift range: 
    $$\nmad = 1.48 \times \rm median(|\Delta z - \rm median(\Delta z)| / (1 + \zspec)),$$
    where $\Delta z = \zspec - \zphot.$ 
\end{itemize}
\begin{itemize}
    \item \textit{Bias}, $b$, measures the mean systematic offset of photometric redshifts versus spectroscopic redshifts of the sample:
    $$b = \left\langle (\zphot - z_{\rm spec}) / (1 + z_{\rm spec}) \right\rangle,$$
    where $\langle \cdot \rangle$ denotes the sample average. 
\end{itemize}
%
\section{\sDS\@ Photometric Redshift Performance Prediction}\label{sec:photoz_prediction_in_7DS}
\begin{figure*}[!ht]
    \centering
    \includegraphics[width = 0.9\textwidth]{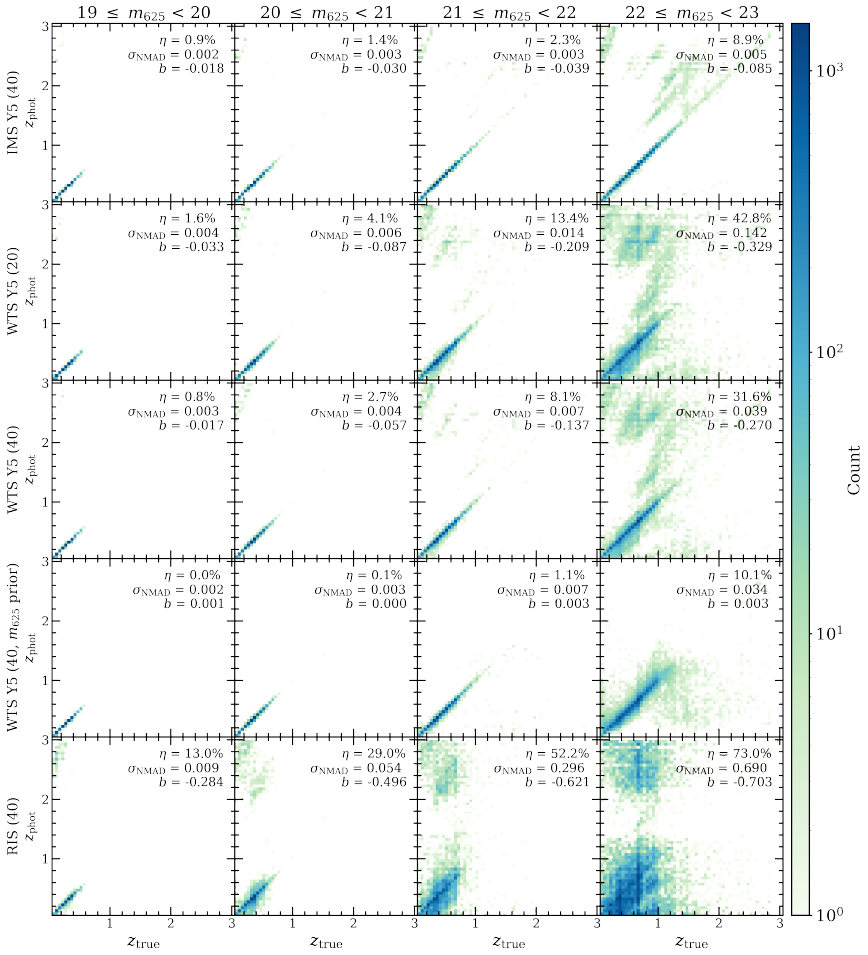}
    \caption{Comparisons between true redshifts $\ztrue$ and predicted photometric redshifts $\zphot$ from the \sDS\@ mock catalog under different input conditions (from top to bottom: template fitted photometric redshifts from \IMS\@ Y5 40 bands, \WTS\@ Y5 20 bands, \WTS\@ Y5 40 bands, \WTS\@ Y5 40 bands with a $m_{625}$ prior, and \RIS\@ 20 bands.)}
    \label{fig:photoz_CF_7DS}
\end{figure*}
In Figure~\ref{fig:photoz_CF_7DS}, the \photoz\@s  for 5 years of \sDS\@ survey under different exposure times and prior conditions are displayed. The impact of increasing exposure time and the number of medium-band filters on \photoz\@s is explored.
\subsection{Photo-z Performance of Each \sDS\@ Survey}
The \IMS\@ Y5 survey (the first row of Figure~\ref{fig:photoz_CF_7DS}) provides highly accurate \photoz\@s with $\nmad$ of about 0.2 to 0.3\% at $m_{625} < 22$. The \photoz\@ accuracy deteriorates a bit at a fainter magnitude bin of $22 \leq m_{625} < 23$, but the survey still provides \photoz\@ accuracy of 0.5\% in $\nmad$. The catastrophic failure and bias is small as well with $\eta < 2.3 \%$ and $b < 0.04$ at $m_{625} < 22$, and $\eta = 8.9 \%$ and $b=-0.085$ at $22 \leq m_{625} < 23$.

The \WTS\@ Y5 survey's performance (from the second to fourth rows of Figure~\ref{fig:photoz_CF_7DS})  is also excellent at brighter magnitudes with $\nmad = 0.002$ to $0.004$ at $m_{625} < 21$ with inclusion of prior improving the \photoz\@ accuracy. For the same magnitude bin, $b = 0.0$ to $0.004$, and $\eta$ is nearly 0\% to a few \% at most. The \photoz\@ accuracy worsens at fainter magnitudes (up to $m_{625} \sim 22$), but is kept at a respectable level of less than a few per cent. But at the faintest magnitudes, both $b$ and $\eta$ become much worse, especially without prior.

The \RIS\@ survey (the last row of Figure~\ref{fig:photoz_CF_7DS}) is the shallowest, so the performance of this survey should be considered only seriously at $m_{625} < 21$. At $19 \leq m_{625} < 20$, which is close to the survey depth, the \photoz\@ accuracy is found to be $\nmad \sim 0.009$, and $\sigma \sim 0.054$ even at $20 \leq m_{625} < 21$. The bias and catastrophic failures are a problem ($\eta > 10 \%$), but an application of a prior should significantly improve these metrics. A more detailed discussion of the impact of priors is elaborated in Section~\ref{subsec:magnitude_prior}.

Overall, our results indicate that \photoz\@ from \sDS\@ can have accuracy better than 1\% and a few tenths \% for brighter sources and out to $z \sim 1$. This level of \photoz\@ performance should allow us to trace large-scale structures of the universe at $z < 1$ and possibly beyond. The implications of these predicted metrics for scientific applications are further discussed in Section~\ref{subsec:comparison-with-previous-medium-band-surveys}.
\subsection{Catastrophic Failures and Survey Depth}\label{subsec:exposure_time}
We investigate the impact of survey depths on $\eta$ by comparing the \WTS\@ Y5 with \RIS\@ and \IMS\@ Y5. This comparison allows us to probe the budget of how the S/N of \sDS\@ influences the detection of spectral features and corresponding \photoz\@s. Overall, increasing the exposure time dramatically reduces the catastrophic failures. This effect is most prominent at a magnitude close to the survey depth. For example, at $21 \leq m_{625} < 22$, \photoz\@s from \RIS\@ with an exposure time of 600 seconds take up 52$\%$ of $\eta$, while those from \IMS\@ Y5 with 156,000 seconds contribute only 2.3$\%$. This trend persists even at the fainter magnitude range $22 \leq m_{625} < 23$. 

For \WTS\@ Y5, most of the failure \photoz\@s at $22 \leq m_{625} < 23$ correspond to galaxies fainter than the limiting magnitude at lower true redshifts $0 < z < 0.5$. Due to the insufficient S/N for the given exposure time, these galaxies are incorrectly identified as fainter galaxies at higher redshifts. However, for \IMS\@ Y5 with 10 times higher exposure time, the failed population is largely recovered ($\eta$ from $ 31.6\%$ to $8.9\%$). Similarly, $\nmad$ also decreases proportionally with $\eta$ across all magnitude bins as exposure time increases. 
\subsection{Effect of the Number of Medium-Band Filters}
To understand the impact of spectral resolution on \photoz\@s, we examine the role of medium-band filter configuration. A unique feature of the \sDS\@ is its overlapping filters (i.e. $m_{400}, m_{413}, \ldots{}, m_{875}, m_{888}$), aiming at the increase in spectral resolution without compromising the S/N compared to the non-overlapping case (i.e. $m_{400}, m_{425}, \ldots{}, m_{850}, m_{875}$). We choose only 20 bands whose filter widths are not overlapping and compare the resulting \photoz\@ with an overlapping 40 band system (the second and third row of Figure~\ref{fig:photoz_CF_7DS}). Compared to \WTS\@ Y5 \photoz\@s results using only 20 bands that are not overlapping, the three metrics--- $\eta$, $\nmad$, and $b$---exhibit better performance for the entire magnitude bins in the 40-band system. This improvement can be seen for both bright and faint samples, highlighting the effectiveness of overlapping filter sets in boosting spectral resolution. The overlapping filters not only compensate for the relatively low S/N associated with the medium-band \FWHM\@ but also provide densely sampled photometric data points. The higher spectral resolution achieved at longer wavelengths for a given \FWHM\@ further aids in the accurate recovery of \photoz\@s for faint, high-redshift sources within the \sDS\@ medium-band system. 

We note that the regions where \photoz\@s deviate from true redshifts largely overlap between the results obtained from the 20- and 40-band systems. While some failure samples are recovered by significantly increasing the exposure time, as in the \IMS\@ Y5 results, uncertainties in determining \photoz\@s still remain. This is partly because degenerate color spaces between low and high redshifts intrinsically continue to hamper correct \SED\@ matching \citep{2003A&A...401...73W, 2010ApJS..189..270C}. We will discuss the capability of the \sDS\@ to capture key spectral features and resolve such degeneracies in more detail in Section~\ref{sec:discussion}. 
\subsection{Magnitude Prior}\label{subsec:magnitude_prior}
The most remarkable improvement after applying the prior (the fourth row of Figure~\ref{fig:photoz_CF_7DS}) is the correction of falsely derived high \photoz\@s ($\zphot > 2$) and low true redshifts ($\ztrue < 1$). The prior breaks the degeneracies between low and high redshifts, showing a consistent trend of converging bias toward zero. However, by design, imposing the prior information is biased toward the lower \photoz\@s. For instance, new failure samples appear at $1.2 < \ztrue <2$ and $\zphot<1$. We suppose that this bias partly arises because this redshift range complies with the longest wavelength coverage of \sDS\@ 888 nm in \sDS\@. Despite this potential problem, the prior should help minimize $\eta$ for brighter galaxies, among which galaxies with $\ztrue > 1.2$ are rare. Inclusion of longer wavelength data from other surveys should help reduce the bias, which will be examined in Section~\ref{sec:synergy_with_other_surveys} and Section~\ref{subsec:discussion2}.
\subsection{Comparison with Previous Medium-Band Surveys}\label{subsec:comparison-with-previous-medium-band-surveys}
We briefly discuss the face-value comparison between the predicted results from \sDS\@ \photoz\@s and other medium-band surveys. Note that the criteria for \photoz\@ metrics---such as catastrophic failure $\eta$ (or outlier fraction), the \photoz\@ calculation code, and wavelength coverage---may differ across surveys. Our prediction reveals a similar or lower $\nmad$ when compared to previous optical medium-band studies such as COMBO-17 (\citealt{2003A&A...401...73W}, $\nmad \approx 0.03$ at $m_{R} < 24$) and MUSYC (\citealt{2010ApJS..189..270C}, $\nmad = 0.008$ at $22 < m_{R} < 23$). On the contrary, our \photoz\@ results indicate higher $\nmad$ and lower $\eta$ than those from narrow-band studies like MiniJPAS (\citealt{2021A&A...654A.101H}, $\nmad < 0.003$ at $m_{R} < 23$ and $odds > 0.82$, $\eta = 0.39$ without any constraints) and PAU (\citealt{2023arXiv231207581N}, see Figure 14). This trend can be explained by the differences in spectral resolution of these different surveys. 

We briefly highlight the scientific applications of our predictions based on the requirements established in the literature. For galaxy evolution studies, the performance metrics achieved with \IMS\@ are highly competitive. For reference,  \citet{Bordoloi2010} and \citet{Hildebrandt2017} suggest that reliable analyses require a minimum scatter of \photoz\@ as $\nmad < 0.05$, while more ambitious surveys such as LSST adopt more stringent thresholds of $\nmad \sim$ 0.01--0.03 \citep{2019ApJ...873..111I}. Because luminosity and stellar mass functions are only moderately sensitive to \photoz\@ errors, our \sDS\@ prediction of $\nmad \leq 0.01$ from \IMS\@ represents a clear advantage. In contrast, studies of galaxy environments demand substantially tighter precision: smearing along the line of sight due to \photoz\@ errors can strongly bias the construction of the density field, further underscoring the value of our results.

As for cosmology, the requirements are considerably more stringent. Precision weak-lensing and large-scale structure measurements rely on minimizing both scatter and systematic bias in redshifts. For example, cosmic shear analyses demand \photoz\@ biases at the level of $b \lesssim 0.002$, with $\eta$ kept to a minimum \citep{knox2006weighing, hearin2010general}. The LSST requirements specify $b < 0.001$ for weak lensing and clustering over its 10-year baseline, together with a scatter of $\nmad < 0.003$ \citep{2019ApJ...873..111I}. The accuracy predicted for \WTS\@ therefore places it within reach of the most demanding cosmological applications.

To sum up, the \sDS\@ surveys occupy a distinctive niche in multi-band surveys between broad- to narrow-band systems. The three surveys, \IMS\@, \WTS\@, and \RIS\@, cover complementary regimes in area and depth, while sharing a consistent medium-band \FWHM\@. Altogether, they form a unified framework that bridges wide, shallow surveys and narrow, deep experiments, thereby maximizing the scientific return of \photoz\@ science across galaxy evolution and cosmology.
\section{Synergy With Other Surveys}\label{sec:synergy_with_other_surveys}
\begin{figure*}[!ht]
    \centering
    \includegraphics[width = 0.9\textwidth]{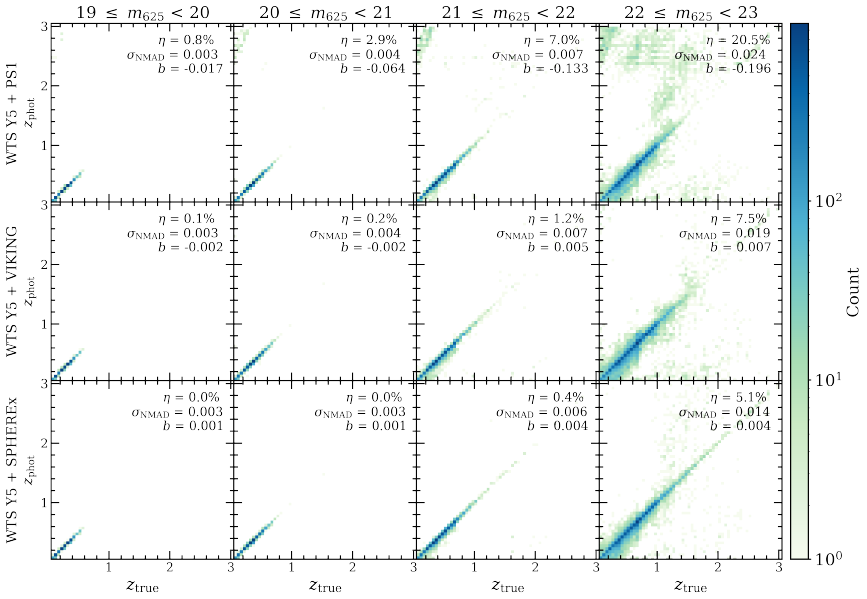}
    \caption{Comparisons between true redshifts and predicted photometric redshifts from the \sDS\@ mock catalog with the combination of other surveys. From top to bottom: photometric redshifts from \WTS\@ Y5 + Pan-STARRS1, \WTS\@ Y5 + VIKING, and \WTS\@ Y5 + SPHEREx all-sky.}
    \label{fig:photoz_CF_7DS_other_surveys}
\end{figure*}
\begin{deluxetable*}{ccccc}[!ht]
\tabletypesize{\scriptsize}
\tablewidth{0pt}
\tablecaption{Prediction results of photometric redshifts in \sDS\@ and addition of other surveys \label{tab:photoz_surveys}}
\tablehead{
\colhead{Surveys} & \colhead{$19 \leq m_{625} < 20$} & \colhead{$20 \leq m_{625} < 21$} & \colhead{$21 \leq m_{625} < 22$} & \colhead{$22 \leq m_{625} < 23$}
}
\startdata
    \WTS\@ Y5 (20) & $(1.6\%, 0.004, -0.033)$ & $(4.1\%, 0.006, -0.087)$ & $(13.4\%, 0.014, -0.209)$ & $(42.8\%, 0.142, -0.329)$\\
    \WTS\@ Y5 (40) & $(0.8\%, 0.003, -0.017)$ & $(2.7\%, 0.004, -0.057)$ & $(8.1\%, 0.007, -0.137)$ & $(31.6\%, 0.039, -0.270)$\\
    \WTS\@ Y5 ($m_{625}$ prior) & $(0.0\%, 0.002, 0.001)$ & $(0.1\%, 0.003, -0.000)$ & $(1.1\%, 0.007, 0.003)$ & $(10.1\%, 0.034, 0.003)$\\
    \RIS\@ Y5 & $(13.0\%, 0.009, -0.284)$  & $(29.0\%, 0.054, -0.496)$  & $(52.2\%, 0.296, -0.621)$  & $(73.0\%, 0.690, -0.703)$ \\
    \IMS\@ Y5 & $(0.9\%, 0.002, -0.018)$  & $(1.4\%, 0.003, -0.030)$  & $(2.3\%, 0.003, -0.039)$  & $(8.9\%, 0.005, -0.085)$ \\
    \WTS\@ Y5 + PS1 & $(0.8\%, 0.003, -0.017)$ & $(2.9\%, 0.004, -0.064)$ & $(7.0\%, 0.007, -0.133)$ & $(20.5\%, 0.024, -0.196)$\\
    \WTS\@ Y5 + VIKING & $(0.1\%, 0.003, -0.002)$ & $(0.2\%, 0.004, -0.002)$ & $(1.2\%, 0.007, 0.005)$ & $(7.5\%, 0.019, 0.007)$\\
    \WTS\@ Y5 + SPHEREx & $(0.0\%, 0.003, 0.001)$ & $(0.0\%, 0.003, 0.001)$ & $(0.4\%, 0.006, 0.004)$ & $(5.1\%, 0.014, 0.004)$\\
\enddata
\tablecomments{($\eta$, $\nmad$, $b$). We use 40 \sDS\@ band filters and template fitting only method except for applying prior (third row).}
\end{deluxetable*}
When combined with other surveys that offer higher S/N and different wavelength coverage, \sDS\@ \photoz\@s can be further improved. In Figure~\ref{fig:photoz_CF_7DS_other_surveys}, we present our findings after integrating data from other surveys (PS1, VIKING, and SPHEREx) with the \sDS\@ dataset. For the \sDS\@ \WTS\@ Y5, we adopt the case with 40 bands and no prior application.
\subsection{\WTS\@ Y5 + PAN-STAARS1}\label{subsubsec:synergy_with_PS1}
By adding PS1 to \sDS\@ (the first row of Figure~\ref{fig:photoz_CF_7DS_other_surveys}), the resulting \photoz\@ metrics improve slightly for samples at $m_{625} < 22$  but by a factor of $\sim2$ for the faintest sample at $m_{625} > 22$ mag, where PS1 presumably complements for the relatively lower S/N of \sDS\@. While samples fainter than \sDS\@ limiting magnitude benefit most from the high S/N broad-band surveys, it is important to note that the location of catastrophic failure samples in the redshift space remains unchanged before and after adding PS1 (e.g., \citealt{2020AJ....159..258G}). As discussed in Section~\ref{subsec:exposure_time}, an increase in the S/N cannot eliminate the contamination caused by intrinsic spectral degeneracies. A more detailed analysis of this issue is presented in the context of \SED\@s in Section~\ref{subsec:discussion1} (see also Figure~\ref{fig:comparison_ps1_7ds_example} for a preview of the spectral resolution of \sDS\@ and PS1).

Furthermore, we confirm that the addition of PS1 broad-bands can sometimes lead to deviations in \photoz\@s due to its high S/N. While 1{,}762 out of 20{,}654 samples brighter than the 5$\sigma$ limiting magnitude $m_{625} \leq 22.62$ benefit from the addition of PS1, 1,018 out of 20{,}654 samples, originally successful with \WTS\@ Y5 alone, show the opposite trend. This occurs because the $\chi^{2}$ values from the template-fitting process are weighted by their errors, and a simplistic approach may result in inaccurate calculations. This finding is consistent with predictions by \citet{2009ApJ...692L...5B} and \citet{2014arXiv1403.5237B}, which suggests that a limited number of broad-band filters can be more prone to color-redshift degeneracies, and the \photoz\@ derived may not benefit from higher S/N to the extent expected. 

Nonetheless, there remains further scope for boosting the combination of medium-band datasets with high S/N broad-band data through more sophisticated approaches. For example, \citet{HernanCaballero2024} demonstrates the effectiveness of conflating the full \zPDF\@ from narrow-band data (J-PAS) and broad-band data (HSC-SSP), rather than relying solely on single-point \photoz\@ estimates. They point out two main challenges in deriving \photoz\@s from both broad- and narrow-band observations: (1) cross-calibration across different data-processing pipelines and instruments, which involves highly complicated systematics that must be controlled to produce homogeneous photometry, and (2) increased vulnerability of faint sources to color degeneracies in template fitting, whereas brighter counterparts are more robust in recovering their true redshifts among a large template set. Therefore, adopting careful strategies that catch the full probability distribution--- encoding spectral features from narrow-band surveys while filtering out spurious peaks secured by high S/N broad-bands---can maximize their synergy \citep{Kovac2010,Barro2019}
\subsection{\WTS\@ Y5 + VIKING}
The extra near-IR information from VIKING (the second row of Figure~\ref{fig:photoz_CF_7DS_other_surveys}) leads to better \photoz\@s compared to \WTS\@ Y5 alone and \WTS\@ Y5 + PS1 across all magnitude bins. The inclusion of near-IR not only enhances the \photoz\@ metrics but also helps recover some of the previously catastrophic failure samples. Specifically, the catastrophic failures, previously populating around low $\ztrue$ and high $\zphot$, are largely mitigated, significantly reducing the bias. This improvement from near-IR coverage differs from the effect of simply increasing S/N. We suppose that 8875~\AA{} to 2.1~$\mu$m wavelength information in VIKING contributes to breaking up the color degeneracies up to redshifts $z = 2$ to 3. Consequently, the clump of points at $\zphot > 2$ and $\ztrue < 1$ is accurately aligned with its true redshift (see the upcoming Section~\ref{subsec:discussion2} for the physical origin of the color degeneracy and Figure~\ref{fig:comparison_NIR_7ds_example} as an example of \SED\@s from \sDS\@ and near-IR surveys). This advantage of incorporating near-IR is consistent with previous works (e.g., \citealt{2009ApJ...692L...5B, 2014arXiv1403.5237B}).
\subsection{\WTS\@ Y5 + SPHEREx}
\begin{figure*}[!ht]
    \centering
    \includegraphics[width = 0.45\textwidth]{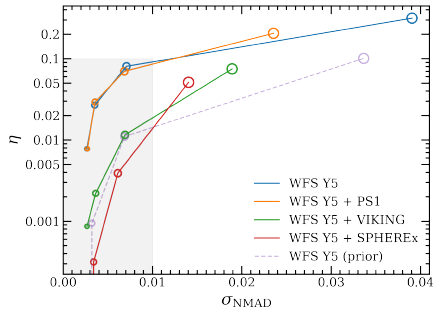}
    \includegraphics[width = 0.45\textwidth]{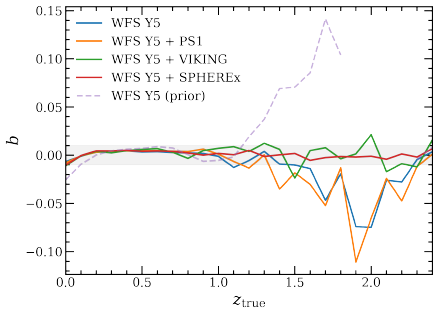}
    \caption{Overall \photoz\@ metrics for \WTS\@ Y5 and combination with other surveys. Left panel: catastrophic failure $\eta$ as a function of NMAD $\nmad$. The size of markers represents the magnitude range (19, 20, 21, 22) in increasing order. Right panel: Bias $b$ as a function of true redshift. The area smaller than the bench marking values, $\nmad = 0.01$, $\eta = 0.10$, and $b = 0.01$, were shaded in gray.}
    \label{fig:photoz_surveys_synergy}
\end{figure*}
A  significant synergy is observed with the combination of \sDS\@ and SPHEREx (the third row of Figure~\ref{fig:photoz_CF_7DS_other_surveys}). SPHEREx extends its wavelength coverage up to 5 $\mu$m, with 96 \LVF\@s in this study. At $19 \leq m_{625} < 22$, $\nmad$ remains sub-percent accuracy, so does $b$ and $\eta$. Even in the fainter magnitude bin $22 \leq m_{625} < 23$, only 2.5$\%$ of samples are classified as catastrophic failures while the \photoz\@ accuracy remains high. It is also notable that the scatter between $\zphot$ and $\ztrue$ is reduced more than in any other combinations probed in our work. 

This improvement further exemplifies the critical role of medium-band filters as a key hunter for spectral features, especially in combination with IR surveys. Previous forecasting studies have already emphasized the mutual benefits of integrating IR and optical broad-band data for \photoz\@ performance (e.g., \citealt{Dore2016,2016arXiv160606374S,Rhodes2017,2018arXiv180505489D,2020AJ....159..258G}). For instance, \citet{2018arXiv180505489D} demonstrates that augmenting SPHEREx-shallow with Euclid data can reduce the $\nmad$ of bright samples at $18 < RIZ < 19$ from 0.013 to 0.005, and for fainter sources near the Euclid magnitude limit ($23.5 < RIZ < 24$) with SPHEREx-deep from 0.054 to 0.015. Similarly, combining LSST and Euclid decreases the standard deviation of \photoz\@ by 20$\%$ at $z < 1$ and lowers the outlier fractions by $40\%$ \citep{2020AJ....159..258G}. Within this context, our results provide strong quantitative evidence that medium-band + IR survey synergies can deliver substantial improvements in redshift accuracy, establishing \sDS\@ as a powerful partner for SPHEREx and future space-based IR missions.

The overall \photoz\@ prediction results in Section~\ref{sec:synergy_with_other_surveys} are illustrated in Figure~\ref{fig:photoz_surveys_synergy} and summarized in Table~\ref{tab:photoz_surveys}.

\section{Discussion}\label{sec:discussion}
As a complementary analysis to the previous statistical approach, we look into individual \SED\@s and evaluate how higher S/N and extra information improve the \photoz\@ metrics in Section~\ref{subsec:discussion1} and Section~\ref{subsec:discussion2}. Additionally, we assess the detectability of emission lines from \sDS\@ in Section~\ref{subsec:discussion3}. \emph{This is the high spot of our study, which shows the power of \sDS\@'s medium-band filter sets.}
\subsection{Broad Band with High S/N vs. Medium Band with High R}\label{subsec:discussion1}
\begin{figure*}[!ht]
    \centering
    \includegraphics[width=0.8\textwidth]{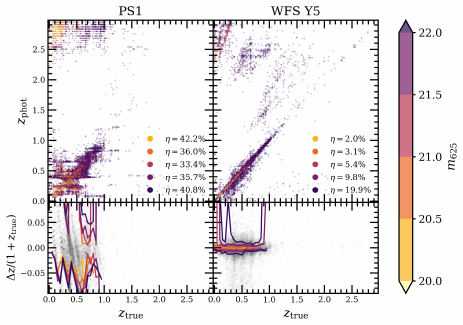}
    \caption{A comparison of \photoz\@s obtained from PS1 and \sDS\@ for a given magnitude bin at $20 \leq m_{625} \leq 22$. No priors are applied for both results.}
    \label{fig:comparison_ps1_7ds}
\end{figure*}
\begin{figure*}[!ht]
    \centering
    \gridline{\fig{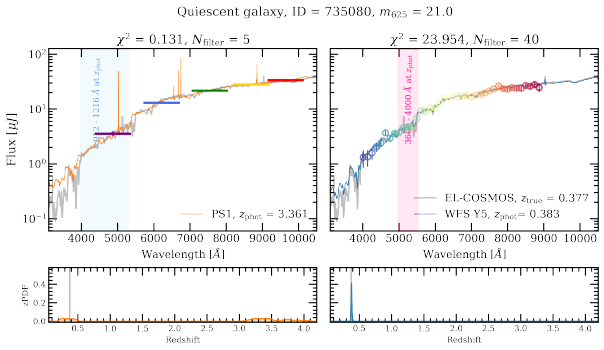}{0.9\textwidth}{(a)}}
    \gridline{\fig{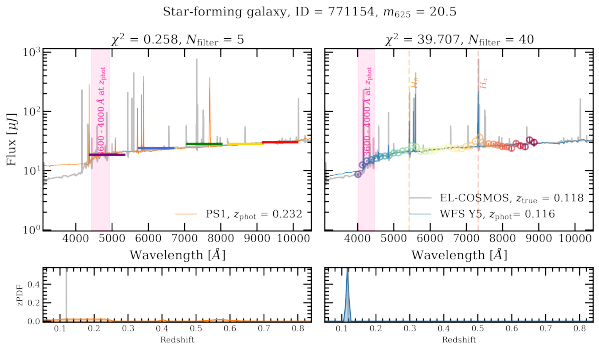}{0.9\textwidth}{(b)}}
    \caption{Examples of matched \SED\@s and \zPDF\@ from \sDS\@ and PS1. (a) A quiescent galaxy at $z_{\rm true} = 0.377$ with a Balmer break. (b) A star-forming galaxy at $z_{\rm true} = 0.118$ with emission lines. The original \SED\@ from EL-COSMOS is shown as a gray line. The orange line is the best-fit \photoz\@ from PS1 and the blue line from \WTS\@ Y5.}
    \label{fig:comparison_ps1_7ds_example}
\end{figure*}
\begin{figure*}[!ht]
    \centering
    \gridline{\fig{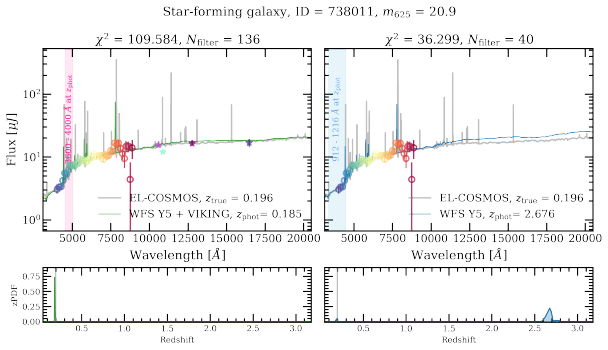}{0.9\textwidth}{(a)}}
    \gridline{\fig{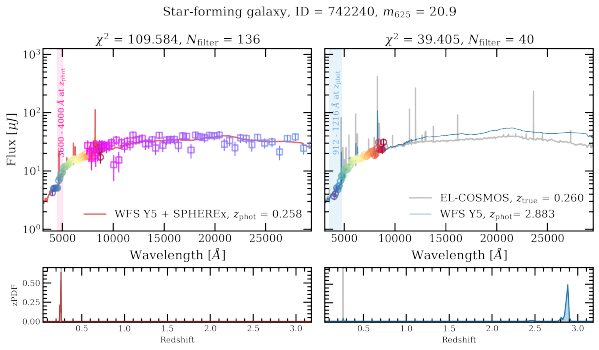}{0.9\textwidth}{(b)}}
    \caption{Examples of matched \SED\@s and \zPDF\@ from \sDS\@ alone and in combination with (a) VIKING and (b) SPHEREx.}
    \label{fig:comparison_NIR_7ds_example}
\end{figure*}
Here, we compare the impact of S/N and spectral resolution on the \photoz\@ accuracy. Figure~\ref{fig:comparison_ps1_7ds} presents a detailed comparison of \photoz\@s obtained from PS1 and \sDS\@ without applying any priors. Overall, $\eta$ for \photoz\@s derived using the template-fitting method in \WTS\@ Y5 increases as sample magnitudes increase, whereas in PS1, $\eta$ remains in a similar range across magnitudes. Despite the deeper 5$\sigma$ depth (23.2) of PS1's $r$-band compared to $m_{625} = 22.62$ in \WTS\@ Y5, \photoz\@s from PS1 can deviate from their true redshifts even for bright samples ($m_{r} < 20$). This is because \photoz\@s from PS1 broad-bands with template fitting methods are more susceptible to color degeneracy, resulting in increased contamination. In contrast, for \sDS\@, most misclassified galaxies are relatively faint, with magnitudes comparable to or fainter than the limiting magnitude. Consistently, as noted in Section~\ref{subsubsec:synergy_with_PS1}, merely increasing S/N does not proportionally improve \photoz\@ accuracy and cannot recover all catastrophic failures. Instead, more detailed spectral diagnostics or additional wavelength information are needed to accurately resolve their features.

Figure~\ref{fig:comparison_ps1_7ds_example} illustrates the comparison of matched \SED\@s from \WTS\@ Y5 and PS1 for a quiescent and a star-forming galaxy. For the quiescent galaxy with a true redshift of $\ztrue = 0.377$, the original \SED\@ from EL-COSMOS clearly shows the Balmer and 4000~\AA{} breaks near the observed wavelength 5000--5500~\AA{}. However, the matched \SED\@ from PS1 misinterprets this feature as a Lyman break for a galaxy at a higher redshift $\zphot = 3.361$. The relatively low spectral resolution of broad-band filters in PS1 fails to distinguish these detailed spectral features, resulting in a much higher \photoz\@. On the other hand, the photometric dataset from \WTS\@ Y5 captures the features before and after the breaks. The \zPDF\@ for the PS1 data shows significant ambiguity, with a broad spread between $0 < \zphot< 0.5$ and $3.0 < \zphot < 3.5$. In comparison, the \zPDF\@ from \sDS\@ is sharply concentrated around the true redshift. This demonstrates that \photoz\@ accuracy benefits highly from densely sampled photo-spectra, which reduces the ambiguity in identifying spectral break features across different redshifts. Lifting such degeneracy is particularly advantageous for quiescent galaxies at low redshift, whose break features are more pronounced than for star-forming galaxies \citep{Stabenau_2008,2019A&A...632A..80G}. 

Emission lines from star-forming galaxies are another crucial factor in determining the \SED\@s. The corresponding example shown in the lower panel of Figure~\ref{fig:comparison_ps1_7ds_example} illustrates this point. For PS1, while the continuum of the matched \SED\@ generally overlaps with the true \SED\@, their emission lines do not. The coarse sampling of the data misses the detailed features of the emission lines, obscuring the accurate \photoz\@ calculation. On the contrary, the \SED\@ from \sDS\@ captures strong emission lines such as $\textrm{H}_{\alpha}$ and $\textrm{H}_{\beta}$ of the true \SED\@. The broad-band system in PS1 dilutes these features when used alone. This result again highlights the importance of spectral resolution in probing spectral features and unscrambling the potential color degeneracies inherent in \SED\@s. 

At last, from the perspective of \zPDF\@, we revisit the possibility of improving the \photoz\@ estimate. As noted in Figure~\ref{fig:comparison_ps1_7ds_example}, the \zPDF\@ derived from PS1, which yields an incorrect \photoz\@, spans much broader redshift space than the case from \sDS\@. This broader distribution makes point estimates such as \texttt{z\_a} and/or \texttt{z\_m2} less representative of the true redshift, thereby degrading \photoz\@ metrics that rely solely on them. Put differently, reducing the width of the \zPDF\@ and an effective strategy would directly lead to more reliable and less contaminated \photoz\@ metrics \citep{CarrascoKind2013,DIsanto2018}.

\subsection{Breaking Color Degeneracies with Addition of near-IR Data}\label{subsec:discussion2}

Along with the need for higher spectral resolution to derive accurate \photoz\@s, solely relying on optical wavelength information faces challenges in differentiating the similar spectral features at low and high redshifts. We analyze the impact of near-IR data on matched \SED\@s, covering from 8875~\AA{} to 2.1~$\mu$m in VIKING and from 0.75 to 5.0~$\mu$m in SPHEREx, as illustrated in Figure~\ref{fig:comparison_NIR_7ds_example}. In the upper panel, the matched \SED\@ from \sDS\@ alone diverges from the true \SED\@ in the near-IR region. Due to the poor S/N near 8875~\AA{} and the absence of longer wavelength information, a low redshift galaxy at $\ztrue = 0.196$ is misclassified as having a higher redshift at $\zphot = 2.676$. However, by providing the VIKING's additional data points covering from 8875~\AA{}{} to 2.1~$\mu$m, the \SED\@ matches the true \SED\@ accurately. A similar improvement is observed with SPHEREx, as shown in the lower panel. We attribute this to SPHEREx's denser and wider coverage from 0.75~$\mu$m to 5.0~$\mu$m, which enables one to sample redshifted 1.6~$\mu$m effectively.
\subsection{Tracing emission lines in \sDS\@ medium-band filters}\label{subsec:discussion3}
\begin{figure*}[!ht]
    \centering
    \includegraphics[width=0.9\textwidth]{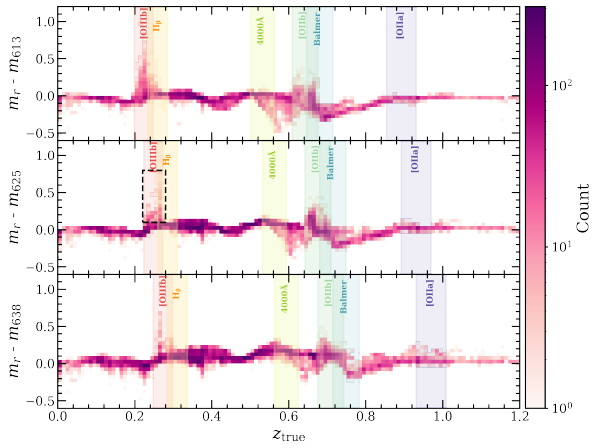}
    \caption{A $m_{r}-m_{625}$ color as a function of true redshifts (middle panel). Adjacent colors of $m_{613}$ (top panel) and $m_{638}$ (bottom panel) are also described.}
    \label{fig:color_excess_6250A}
\end{figure*}
Thanks to the 25~nm \FWHM\@ of \sDS\@ medium-band filters, it is possible to detect strong emission lines indirectly. The emission line diagnostics provide fundamental constraints on galactic properties, linked to the ionized interstellar medium  \citep{Kewley2019}. For example, from diagnostics such as electron density and ISM pressure, one can infer the ionization state, chemical abundance, and the dominant excitation sources of galaxies (e.g. \citealt{Kewley2013, Peimbert2017}). The use of photo-spectra for diagnosing emission features has been suggested (e.g. \citealt{2019MNRAS.486.5104L}) and implemented in various studies (e.g. \citealt{2021MNRAS.508.3860G, 2022A&A...665A..95I, 2023ApJS..268...64W, 2023ApJ...958L..14W, 2024ApJ...967L..17S}). For example, \citet{2021A&A...653A..31B} demonstrated that narrow-band filters with a 145~\AA{} width can detect emission lines from star-forming galaxies, thereby improving the accuracy of \photoz\@ measurements. Moreover, \citet{2022MNRAS.515..146R} also suggests the direct measurement of 4000~\AA{} break using narrow-band filters, which allows for the reproduction of empirical relations such as D4000-SFR and D4000-mass relation in spectroscopy.

In this regard, we can experiment to check \sDS\@'s potential for tracing emission lines by comparing the color excess between the broad-band and medium-band filters. An example of $m_{r}-m_{625}$ as a function of redshifts is shown in Figure~\ref{fig:color_excess_6250A}. Interestingly, this comparison reveals that the color excess between broad- and medium-band filters corresponds to known emission lines and breaks. Specifically, redshifted emission lines (e.g. $\textrm{H}_{\alpha}$, $\textrm{H}_{\beta}$, [OIII], and [OII]) coincide to lie at the positive $m_{r}-m_{625}$ color while spectral breaks (4000~\AA{} and Balmer break) are associated with negative values. This alignment suggests that \sDS\@ can effectively trace these important spectral features. Other examples of HSC $g$ and $i$ band color excess are described in Figure~\ref{fig:color_excess_4750A} and Figure~\ref{fig:color_excess_7750A}.

We select galaxies exhibiting strong color excess and control samples to investigate the relationship between the strength of this excess and the potential to catch spectral features for \photoz\@ calculations. We focus on sample galaxies at $0.2 < \ztrue < 0.3$ and $m_{r}-m_{625} > 0.2$ with S/N $>$ 10 at $m_{625}$ (highlighted in the dashed box in the second row of Figure~\ref{fig:color_excess_6250A}) as potential strong emission line candidates. We compare these candidates with residual samples that meet the same S/N criteria but do not show a significant color excess ($0.2 < \ztrue < 0.3$, $m_{r}-m_{625} < 0.2$, and S/N $>$ 10 at $m_{625}$). The color excess threshold of 0.2 mag is set to correspond to a 5$\sigma$ uncertainty at the limiting magnitude of $m_{625}$. In total, we identify 54 galaxies with $m_{r} - m_{625} > 0.2$ and 1,740 with $m_{r} - m_{625} < 0.2$. For more precise \photoz\@ estimates, the ability to detect emission lines is also beneficial. While no catastrophic failure galaxies are found with color excess samples, 187 galaxies ($\sim 11\%$) without color excess deviate by more than $15\%$ from their true redshifts. 

We confirm a similar excess pattern when replacing the HSC broad-band with fluxes near the given filter. To summarize, \sDS\@ photometric data can serve as a useful indicator of emission lines.

\section{Conclusion}\label{sec:conclusion}
In this paper, we explore how reliably medium-band filters in the 7-Dimensional Sky Survey produce photometric redshifts. We use simulation data that accounts for observational systematics and survey plans to provide a comprehensive prediction of photometric redshifts in \sDS\@. Our findings indicate that further improvement is achievable when combined with other surveys, SPHEREx in particular. We conclude our findings below. 

\begin{enumerate}
\item Using model \SED\@s from EL-COSMOS, we construct mock photometric data that resemble the \sDS\@ setup. This includes accounting for telescope and detector efficiencies, as well as sky transmission at the observatory site in Chile. 
We produce a mock catalog containing half a million sources and derive photometric redshifts using the \texttt{EAZY} code, covering the 5-year Wide-field Time-Domain Survey (\WTS\@ Y5), Reference Imaging Survey (\RIS\@), and 5-year Intensive Monitoring Survey (\IMS\@ Y5) in \sDS\@, along with other surveys including Pan-STARRS1, VIKING, and SPHEREx all-sky.

\item Overall, the \WTS\@ Y5 observed with 40 medium-band filters achieves a catastrophic failure fraction $\eta$ from 0.8 to $8.1\%$, NMAD $\nmad$ from 0.003 to 0.007, and bias $b$ from $-0.017$ to $-0.137$ at $19 \leq m_{625} < 22$. The \IMS\@ Y5, whose on-source exposure time is 10 times longer than the \WTS\@ Y5, can significantly reduce catastrophic failures at faint magnitudes around the \WTS\@ Y5 limiting magnitudes. However, merely increasing exposure time does not fully resolve contamination issues. Instead, additional information, such as magnitude priors or near-IR coverage, is crucial for further improvements. 

\item Combining \sDS\@ with other surveys PS1, VIKING, and SPHEREx all-sky yields overall improvements of \photoz\@s. While PS1 helps recover true redshifts from samples with insufficient S/N, it does not entirely resolve the intrinsic color degeneracies between faint galaxies at low and high redshifts. The association with near-IR information from VIKING and SPHERE, however, plays a complementary role by covering the wavelengths not probed by \sDS\@. 

\item We compare the matched \SED\@s of individual galaxies obtained from template fitting. The \sDS\@ data effectively capture spectral features of 4000~\AA{}{} and Balmer break for quiescent galaxies and emission lines for star-forming galaxies, despite having lower S/N than PS1. Furthermore, the optical-only \sDS\@ data can obtain more constraining power in synergy with near-IR data from VIKING and SPHEREx. They provide the coverage of  8875~\AA{} to 2.1~$\mu$m from VIKING and 0.75--5.0~$\mu$m from SPHEREx,  respectively, as \zPDF\@ becomes sharper and concentrated around the true redshift. 

\item We demonstrate the potential application of medium-band filters to trace strong emission lines. By analyzing the color excess between \sDS\@ medium-band and adjacent PS1 broad-band photometric data, we identify features corresponding to spectral breaks and emission lines. 
\end{enumerate}

To conclude, \sDS\@ offers promising prospects for accurate redshift measurements. The medium-band filters, with a spectral resolution of $R \sim 50$ of \FWHM\@ = 25~nm, play a crucial role in detecting spectral diagnostics that are otherwise missed. With the aid of additional data, particularly from SPHEREx, we expect to extend these capabilities further to enhance our understanding of astrophysics and cosmology. 

\begin{acknowledgments}
This work was supported by the National Research Foundation of Korea (NRF) grant No. 2021M3F7A1084525, funded by the Korea government (MSIT). SL acknowledges support from the National Research Foundation of Korea (NRF) grant (RS-2025-00573214) funded by the Korea government(MSIT). J.H.K. acknowledges the support from the National Research Foundation of Korea (NRF) grants, No. 2021M3F7A1084525 and No. 2020R1A2C3011091, and the Institute of Information \& Communications Technology Planning \& Evaluation (IITP) grant, No. RS-2021-II212068 funded by the Korean government (MSIT).

We acknowledge the use of EL-COSMOS. This research has made use of the ASPIC database, operated at CeSAM/LAM, Marseille, France. 
\end{acknowledgments}
\clearpage
\appendix
\counterwithin{figure}{section}
\counterwithin{table}{section}
\section{\sDS\@ data specification}
\begin{table*}[!h]
\centering
\caption{Data Specifications of 7-Dimensional Telescope, detector, and atmospheric transmission}
\label{tab:data_specification}
\begin{tabular}[t]{lc}
\toprule
&Telescopes\\
\midrule
Telescope&Planewave DR500\\
Number of telescopes&20\\
Aperture size& 50.8 cm ($f/3$)\\
Effective focal length & 1537.3 mm\\
Central Obscuration & 29.8 cm\\ 
\midrule
& Detector\\
\midrule
Camera& Moravian Camera C3-61000PRO\\
Detector & IMX455 rolling shutter back-illuminated CMOS\\
Resolution & $9576 \times 6388$\\
Pixel size & 3.76 $\mu$m $\times$ 3.76 $\mu$m\\
Dark current & 0.01 \textrm{$-$e/s}\\
Readout noise & 3 \textrm{$-$e}\\
\midrule
&Model for Paranal site (VLT)\\
\midrule
Airmass & 1.3\\
PWV & 2.5 mm\\
\bottomrule
\end{tabular}
\end{table*}
\clearpage
\section{Color Excess}
\begin{figure*}[!ht]
    \centering
    \includegraphics[width=0.9\textwidth]{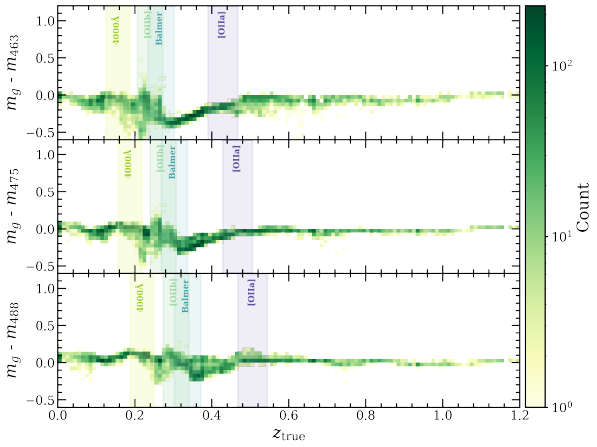}
    \caption{Same as Figure~\ref{fig:color_excess_6250A} but for $m_{g}-m_{475}$ color.}
    \label{fig:color_excess_4750A}
\end{figure*}
\begin{figure*}[!ht]
    \centering
    \includegraphics[width=0.9\textwidth]{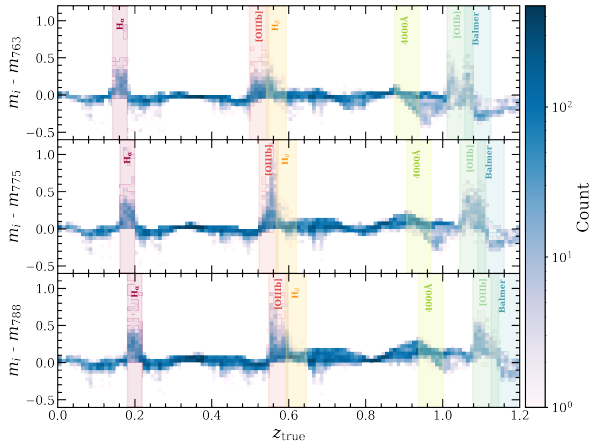}
    \caption{Same as Figure~\ref{fig:color_excess_6250A} but for $m_{i}-m_{775}$ color.}
    \label{fig:color_excess_7750A}
\end{figure*}
\clearpage
\bibliography{references}
\bibliographystyle{aasjournal}
\end{CJK*}
\end{document}